\documentclass[12pt]{article}

\usepackage{epsfig, bm, amsfonts, amssymb}

\DeclareMathAlphabet{\mathitbf}{T1}{cmr}{bx}{it} 

 \usepackage{graphicx}
\RequirePackage{color}

 \definecolor{MyDarkGreen}{rgb}{0.02,0.60,0.06}

\title{\bf Scaling Behavior of the Heisenberg Model in Three Dimensions}
\author{ 
{\it A. Gordillo-Guerrero$^{\,1,2}$,} {\it R.~Kenna$^{\,3}$} and {\it J.J. Ruiz-Lorenzo$^{\,4,2}$}\\~\\
$^1$ Departamento de Ingenier\'{\i}a El\'ectrica, Electr\'onica y Autom\'atica,\\
 Universidad de Extremadura,
Avda Universidad s/n, \\
C\'aceres, 10071, Spain. 
{}\\~\\
$^2$ Instituto de Biocomputaci\'on and\\
  F\'{\i}sica de Sistemas Complejos (BIFI),\\
Zaragoza, 50009, Spain.
{}\\~\\
$^3$ Applied Mathematics Research Centre,\\
Coventry University,
Coventry, CV1 5FB, England
{}\\~\\
$^4$ Departamento de F\'{\i}sica,\\
 Universidad de Extremadura,
Avda Elvas s/n, \\
Badajoz, 06071, Spain. 
{}\\~\\
}
\begin{document}
\maketitle
{\Large
  \begin{abstract}
%
We report on extensive numerical simulations of the three-dimensional
Heisenberg model and its analysis through finite-size scaling of Lee-Yang
zeros.  Besides the critical regime, we also investigate scaling in the
ferromagnetic phase.  We show that, in this case of broken symmetry, the
corrections to scaling contain information on the Goldstone modes.  We present
a comprehensive Lee-Yang analysis, including the density of zeros and confirm
recent numerical estimates for critical exponents.
                        \end{abstract} }
%
  \thispagestyle{empty}
%
%
  \newpage
%
                  \pagenumbering{arabic}

\section{Introduction}
\label{I}
\setcounter{equation}{0}

The universality concept is commonly stated together with the hypotheses of
scaling and finite-size scaling for thermodynamic functions close to the
critical points associated with continuous phase transitions.  The theory of
finite-size scaling has been mostly successful in determining critical and
non-critical properties of bulk systems from those of their finite or
partially finite counterparts \cite{O03}.  Although comparisons between theory
and experiment, as well as between the variety of theoretical approaches,
yield good overall agreement in the main, difficulties in reconciling details
of scaling in a number of important systems remain \cite{PhFr10,DhDh12}.
These include some systems with continuous symmetry groups, such as the those
in the three-dimensional $O(3)$ Heisenberg universality class.  Experimental
realizations of this model include isotropic ferromagnets with and without
quenched disorder, e.g. Ni, EuO and La$_{1-x}$A$_x$MnO$_3$.  Precise
theoretical estimates for the critical temperature and critical exponents are
contained in Refs.\cite{CaHa02,UexO3_07} for the pure and site-diluted
Heisenberg models with quenched disorder.  A review of theoretical and
experimental measurements of critical exponents for the Heisenberg model is
contained in Ref.~\cite{PeVi02}.

Here we study the Heisenberg model in three dimensions.  Our objective is not
to revisit old ground but to investigate, for the first time, the Lee-Yang
zeros of this continuous-symmetry-group model through Monte Carlo simulations
\cite{LEEYANG}.  We do this through finite-size scaling, primarily at the
critical point, but also in the ferromagnetic regime.  One motivation is to
investigate the Goldstone modes in the broken phase, which affect the
corrections to scaling there. The symmetric phase manifests the Yang-Lee edge,
which also has not previously been analyzed numerically in this model. We also
investigate the crossover in behavior of the density of zeros from the
critical point to the ferromagnetic phase.

In the next section we outline the Heisenberg model and briefly discuss the
observables we focus on in this paper.  In Section~3 we give details of the
Monte Carlo simulations.  The outcomes of the simulations are analyzed in
Section~4.  A compact scaling description in terms of densities of zeros is
given in Section~5 and we conclude in Section~6.

\section{Model and observables}
\label{M}
\setcounter{equation}{0}

The Heisenberg model in $d$ dimensions may be defined in terms of $O(3)$ spin
variables placed at the nodes of a cubic lattice.  The volume of the lattice
is $V=L^d$ where $L$ is its linear size and the lattice constant has been set
to one.  The model is governed by a Hamiltonian $\mathcal{H}$ given by
\begin{equation}
\beta \mathcal{H}= 
-\beta\sum_{<i,j>} \boldsymbol{\mathit{S}_i}\cdot\boldsymbol{\mathit{S}_j}
-\boldsymbol{h}\cdot \sum_{i} \boldsymbol{\mathit{S}_i} .
\label{heismodel}
\end{equation}
Here $\beta = 1/(k_BT)$, where $T$ is the temperature and $\boldsymbol{h}$ is
an external magnetic field.  The $\boldsymbol{\mathit{S}_i}$ are
three-dimensional vectors of unit modulus and the first sum is extended only
over nearest neighbors. We henceforth set the Boltzmann constant $k_B$ to
unity.  We define the total nearest-neighbor energy as
\begin{equation}
E = - \sum_{\langle
i,j\rangle}\boldsymbol{\mathit{S}_i} \cdot \boldsymbol{\mathit{S}_j} 
\label{totalenergy}
\end{equation}
and the total magnetization density as a three-component vector
\begin{equation}
\boldsymbol{\mathit{M}} =(M_x, M_y, M_z)=\sum_i \boldsymbol{\mathit{S}_i}.
\label{vectorialmag}
\end{equation}
The partition function is 
\begin{equation}
Z_L(T,h) =\sum_{\{\boldsymbol{\mathit{S}_i}\}} 
 \exp{\left({-\beta \mathcal{H}}\right)} 
= \sum_{\{\boldsymbol{\mathit{S}_i}\}} 
  \exp{
       \left({-\beta E + \boldsymbol{\mathit{h}}  \cdot \boldsymbol{\mathit{M}}}\right)
       }.
\label{Zdefn}
\end{equation}
The  susceptibility is defined through the derivatives
\begin{eqnarray}
{\mathitbf{\nabla}}_h  \ln{Z}& =& \frac{1}{V}\left\langle {\boldsymbol{\mathit{M}}} \right\rangle,\\ 
 \chi_L  (T,h)
  &=& \frac{1}{V} \mathitbf{\nabla_h}^2\ln{Z_L(T,h)}
  = \frac{1}{V}\left( \left\langle  \boldsymbol{\mathit{M}}^2 \right\rangle -
 \left\langle {\boldsymbol{\mathit{M}}} \right\rangle^2 \right),
 \label{childerv}
\end{eqnarray}
in which the thermal average is denoted by $\langle \cdots \rangle$.  Because
the probability of reaching every minimal value for the free energy is
non-vanishing, the thermal average of Eq.~(\ref{vectorialmag}) is zero in the
absence of an external field, for a finite lattice.  While this is an accurate
finite-size counterpart for the susceptibility in the symmetric phase, it
cannot be used to capture the connected susceptibility in the broken phase.
Therefore we have to introduce separate definitions for the connected and
non-connected finite-size susceptibilities, namely
\begin{equation}
\tilde{\chi}_L (T,h=0)
=
\frac{1}{V}\left( \left\langle  \boldsymbol{\mathit{M}}^2 \right\rangle -
 \left\langle |\boldsymbol{\mathit{M}}| \right\rangle^2 \right),
\label{susceptibility}
\end{equation}
and
\begin{equation}
\chi^{\rm{(nc})}_L (T,h=0) =\frac{1}{V} \left\langle  \boldsymbol{\mathit{M}}^2 \right\rangle .
\label{susceptibilitync}
\end{equation}
For numerical measurements on a finite lattice it is appropriate to use
$\tilde{\chi}_L$ and ${\chi}^{\rm{(nc})}_L$ in the ferromagnetic and
paramagnetic regimes, respectively.  One should not use ~$\tilde{\chi}_L $ in
the paramagnetic phase because, unlike $\langle \boldsymbol{\mathit{M}}
\rangle$, $\langle |\boldsymbol{\mathit{M}}| \rangle$ does not vanish there
for finite-size systems.  Indeed, the usage of $\langle
|\boldsymbol{\mathit{M}}| \rangle$ in the symmetric phase would be tantamount
to the introduction of a non-vanishing external field there.  There is no
order parameter for finite-size systems (because they do not manifest a phase
transition), but $\chi^{\rm{(nc})}_L$ and $\tilde{\chi}_L$ each approach
$\chi_\infty$ in the thermodynamic limit.

Inspired by the fundamental theorem of algebra, Lee and Yang introduced the
idea of complex zeros of the partition function as a way to investigate the
onset and properties of phase transitions \cite{LEEYANG}.  The resulting
approach constitutes a fundamental theory of phase transitions \cite{Wu}.  In
the paramagnetic phase, the Lee-Yang zeros in the complex $h$-plane remain
away from the real magnetic-field axis, as proved in Ref.\cite{GaMSRo67}.
This means there exists a gap on the imaginary $h$-axis in which the density
of zeros is zero.  The free energy is analytic in $h$ in the gap and no phase
transition can occur as a function of $h$.  The ends of the non-vanishing
distribution of zeros was termed the Yang-Lee edge in Ref.~\cite{Fi78}.  The
proof that the Lee-Yang zeros of the partition function are located on the
imaginary $h$-axis for the Heisenberg ferromagnet was given in
Ref.~\cite{Asano}.  Here we present a numerical investigation into the
Lee-Yang zeros for the model in three dimensions.

Following Ref.\cite{JJRu97}, for example, in order to find the Lee-Yang (LY)
zeros of the system we write the partition function in an imaginary field of
strength $ir$ as
\begin{equation}
Z_L(T,ir) =  Z_L(T,0) \langle{ \cos(rM) +i \sin(rM) }\rangle  ,
\label{Z_imaginary}
\end{equation}
where $M$ is the component of $\boldsymbol{\mathit{M}}$ picked out by the
field.  Here the thermal average is a real measure, taken with $Z(T,h=0)$.
Since the mean value of an observable which is not invariant under $O(3)$ is
automatically zero, odd moments of the magnetization vanish.  Therefore the
partition function in a pure imaginary magnetic field is real and the zeros
are given by $\langle \cos (rM) \rangle = 0$.  In the ferromagnetic phase, and
in analogy with Eq.(\ref{susceptibility}), one may use for $M$ the modulus
$|\boldsymbol{\mathit{M}}|$.  Then one seeks the zeros as solutions
to\begin{equation} \langle \cos (r|\boldsymbol{\mathit{M}}|) \rangle = 0 .
\label{cos_LY}
\end{equation}
However, just as Eq.(\ref{susceptibility}) is inappropriate in the
high-temperature phase, so is Eq.(\ref{cos_LY}) unsuitable there. Instead, and
in analogy to Eq.(\ref{susceptibilitync}), one has to use an explicit
component for $M$, say $M=M_x$.
\begin{equation}
\langle \cos (rM_x) \rangle = 0 .
\label{cos_LYx}
\end{equation}
In this way we can attempt to obtain the zeros of the partition function for
each $L$ in the various regimes.  The resulting $j$th, temperature-dependent,
Lee-Yang zero is denoted by $r_j(T;L)$, the zero with $j=1$ being the
smallest.

\subsection{Scaling of the thermodynamic functions and zeros}

In the limit of infinite volume, the standard expressions for the leading
thermal scaling are $ \chi_\infty(T) \sim |t|^{-\gamma}$, $ m_\infty(h) \sim
h^{{1}/{\delta}}$, $ \xi_\infty (T) \sim |t|^{-\nu}$ and $ r_1(T) \sim
t^{\Delta}$, for~$t>0$.  Here, $t$ is the reduced temperature, $(T-T_c)/T_c$,
and we suppress writing explicit dependency on $h$ or $t$ when they vanish.
In the following, we focus on the finite-size scaling (FSS) of the
susceptibility and zeros.  According to standard FSS theory, these are
obtained through the substitution $\xi_\infty(T) \rightarrow \xi_L(T_c) \sim
L$ or $t \rightarrow L^{-1/\nu}$.  Therefore we expect to extract the leading
scaling behavior through
\begin{equation}
  \chi_L(T_c) \sim   L^{\frac{\gamma}{\nu}}, \quad 
 {r_j}(T_c;L)   \sim   L^{-\frac{\Delta}{\nu}}.
 \label{rjL}
\end{equation}
We can  estimate $\gamma / \nu$   and $\Delta/\nu$ through the scaling relations 
\begin{equation}
\frac{\gamma }{ \nu}  = 2 - \eta, \quad {\mbox{and}} \quad
\frac{\Delta}{\nu} = \frac{d+2-\eta}{2},
\end{equation} 
provided the anomalous dimension $\eta$ is known.  The most recent measurement
of this critical exponent is $\eta=0.0391(9)$~\cite{UexO3_07} and from this
one obtains
\begin{equation}
\frac{\gamma }{ \nu}  = 1.609(9) \quad {\mbox{and}} \quad
\frac{\Delta}{\nu} = 2.4805(4).
\end{equation} 
These values will be used throughout this work.

\subsection{Compact scaling of Lee-Yang zeros}
\label{compact}

In Ref.\cite{IPZ83}, it was suggested that the partition function zeros could
scale, in the critical region, as a fraction of the total number of zeros,
i.e., as a function of $j/L^d$, for large values of the index $j$.  In fact
many models give scaling in the ratio $(j-\epsilon)/L^d$ in which $\epsilon =
1/2$ \cite{JaKe2001,Janus}.  If such a functional form is widespread for Lee-Yang
zeros, it suggests that Eq.(\ref{rjL}) be promoted to the more comprehensive
form
\begin{equation}
 {r_j}(T;L) \sim   \left({\frac{j-\epsilon}{L^d}}\right)^{C(T)},
\label{epsilon}
\end{equation}
for $T \le T_c$.  To investigate this further, first write the finite-size
partition function in terms of its Lee-Yang zeros as
\begin{equation}
 Z_L(T, h) =  A\prod_{j=1}^{V/2}{(h - h_j)(h-h_j^*)},
\label{a1}
\end{equation}
where $A$ is non-vanishing (as a function of $h$) and $*$ means complex conjugation.
Then the susceptibility is
\begin{equation}
 \chi_L(T)  
            =  -\frac{1}{V}\sum_{j=1}^{V/2}{\left({\frac{1}{h_j^2} + \frac{1}{{h_j^*}^2}}\right)}.
 \label{a2}
\end{equation}
When the Lee-Yang theorem \cite{LEEYANG} holds and $h_j = i r_j$ is purely
imaginary this gives that
\begin{equation}
 \chi_L(T) = \frac{2}{V}\sum_{j=1}^{V/2}{\frac{1}{r_j^2(T;L)}}.
\label{a3}
\end{equation}

Eq.(\ref{a3}) relates the Lee-Yang zeros to the susceptibility defined through
Eq.(\ref{childerv}), i.e., through the second derivative of the partition
function.  To gain insight into the behavior of the zeros away from
criticality, we consider the $T \rightarrow 0$ and $T \rightarrow \infty$
limits of this susceptibility.  For a finite-size system, the full
susceptibility $\chi_L$ from Eq.(\ref{childerv}) coincides with the
non-connected version ${\chi}_L^{\rm{(nc)}}$ defined through
Eq.(\ref{susceptibilitync}), and at low temperatures, $\chi_L(T\ll 1) =
{\chi}_L^{\rm{(nc)}}(T\ll 1) \simeq  V$.  If $T \rightarrow \infty$, on the other hand,
$\chi_L(T \rightarrow \infty) = {\chi}_L^{\rm{(nc)}}(T \rightarrow \infty) =
1$.  Note that, these are different to the modified susceptibility
$\tilde{\chi}_L(T)$ as defined in Eq.(\ref{susceptibility}).  This is only
used as a replacement for $\chi_L(T)$ below $T_c$ and, in the low-temperature
limit where the spins align, it is $\tilde{\chi}_L (T=0) = 0$.  However, even
in the paramagnetic phase, it is the susceptibility $\chi_L$ (equivalently
${\chi}_L^{\rm{(nc)}}$), and not $\tilde{\chi}_L$ that is related to the
Lee-Yang zeros through Eq.(\ref{a3}).

In the absence of a YL edge, when the zeros can pinch the real axis, we assume
the form (\ref{epsilon}) for the comprehensive scaling of the zeros.
Inserting into Eq.(\ref{a3}), one finds that the leading finite-size behavior
of the susceptibility is
\begin{equation}
 \chi_L(T)
 \sim
    L^{d(2C-1)} ,
\end{equation}
which comes from  the contributions from the lowest zeros.

This recovers the FSS formula (\ref{rjL}) in the critical regime provided
$C(T_c)={\Delta}/{\nu d}$.  It also recovers the scaling $\chi_L(T<T_c) \sim
L^d$ in the ferromagnetic phase if $C=1$ there.  The ansatz (\ref{epsilon})
fails in the paramagnetic phase, however, because, it does not take into
account the Yang-Lee edge and, plugged into Eq.(\ref{a3}), it would lead to a
spurious logarithmic divergence in the susceptibility there.

\subsection{Corrections to scaling}

At the critical point, the corrections to leading finite-size scaling are
governed by the $\omega$ exponent.  For the susceptibility and Lee-Yang zeros,
one expects
\begin{eqnarray}
  \chi_L(T_c) & \sim &  L^{\frac{\gamma}{\nu}}\left[{1 + \mathcal{O}(L^{-\omega})}\right], 
  \label{FSSofchirr} \\
 {r_j}(T_c;L)& \sim &  L^{-\frac{\Delta}{\nu}}\left[{1 + \mathcal{O}(L^{-\omega})}\right].
 \label{rjLerr}
\end{eqnarray}
The widely accepted measured value for the correction exponent is $\omega
\approx 0.8$ \cite{UexO3_07,PeVi02}.

Away from the critical temperature one may also expect corrections to scaling.
Since the model under consideration has continuous symmetry group, the effects
of Goldstone modes in the ferromagnetic regime are of interest (the
ferromagnetic phase is also critical).  Since these modes are massless, the
corresponding propagator is $1/p^2$ in momentum space, producing an $L^2$
divergence in the connected susceptibility.  The longitudinal susceptibility
on the other hand, diverges as $1/p^{4-d} = 1/p$, inducing a correction
proportional to $L$.  Therefore, the susceptibility for the Heisenberg model
in the ferromagnetic phase (in the presence of Goldstone modes) may be
expected to scale as
\begin{equation}
 \chi_L (T<T_c) \sim L^3 \left[{ 1+ \mathcal{O}(L^{-1}) + \mathcal{O}(L^{-2})}\right].
 \label{chibelow}
\end{equation}
In the Ising case, on the other hand, the absence of Goldstone modes suggests
the absence of such correction.  There, one expects Eq.(\ref{chibelow}) to be
replaced by $\chi_L (T<T_c) \sim L^3 \left[{ 1+ \mathcal{O}(L^{-3})}\right]$.
Therefore the corrections to scaling in the broken phase deliver information
on the existence of Goldstone modes.  As we have seen, the Lee-Yang zeros are
closely related to the susceptibility.  They may therefore be expected to
carry the same correction-to-scaling behavior.  One then expects, for the
zeros in the low-temperature phase,
\begin{equation}
 r_j(T<T_c; L)  
 \sim   
 L^{-d}\left[{1 + \mathcal{O}(L^{-1})}\right].
 \label{LYGB}
\end{equation}

Combining with the index-dependency suggested in the previous subsection, one
expects a comprehensive scaling behavior for the Lee-Yang zeros for the
Heisenberg model as
\begin{equation}
 r_j(T;L) =   \left({\frac{j-\epsilon}{L^d}}\right)^{C} \left[{1+{\mathcal{O}}(L^{-E})}\right],
\label{a77}
\end{equation}
with $C=\Delta/\nu d$~ and $E=\omega$ when  $T \approx T_c$, and $C=1$ and $E=1$ when $T \ll T_c$.

\section{Simulation details}
\label{S}
\setcounter{equation}{0}

\begin{figure}[!t]
\begin{center}
\includegraphics[width=0.45\columnwidth, angle=270]{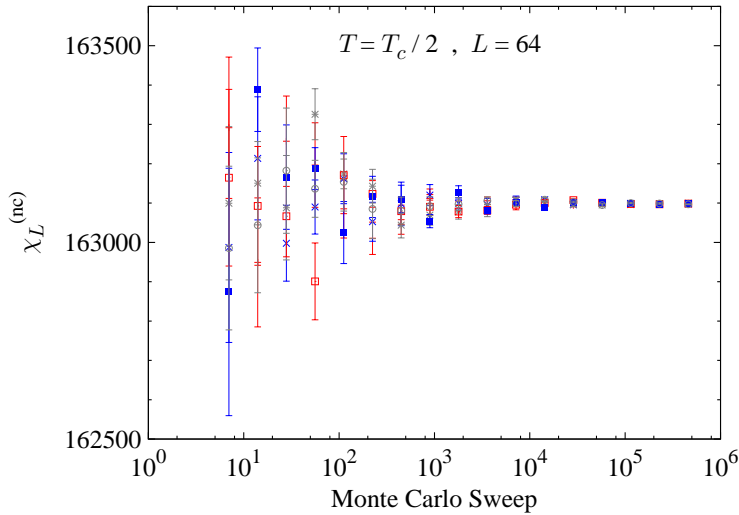}
\caption{Log-binning of susceptibility for five random pseudosamples (in color
  online) with $L=64$ simulated at $T=T_c/2$. Error bars are typical
  deviations in each bin. The first block only includes seven measurements,
  explaining the deviations for small times.}
\label{thermatest}
\end{center}
\end{figure}

\begin{table}[b]
\begin{center}
\begin{tabular}{|l|r|r|r|r|r|r|r|r|} \hline \hline
 $L               $ &   8       &  12 & 16 & 24 & 32 & 48 & 64   \\
 $N_\mathrm{Wolff}$ &  10      &  20 & 25 & 40 & 50 & 75 & 100  \\
\hline \hline
\end{tabular}
\caption{The Monte Carlo sweep size for each system size $L$.  Here,
  $N_\mathrm{Wolff}$ denotes the number of Wolff updates performed after the
  partial Metropolis update.}
\label{tab1}
\end{center}
\end{table}

We simulated the Heisenberg model on regular, cubic lattices with linear sizes
$L=8,12,16,24,32,48$ and $64$.  We used periodic boundary conditions and in
each case 20 pseudosamples were used to average out the thermal noise.  We
performed our simulations at several different values of the system
temperature.  We use the estimate for the critical temperature
$\beta_c=1/T_c=0.693$ from Ref.~\cite{UCMOND3}.  Apart from this value, we
also simulated at two lower temperatures: $T=2T_c/3$ ($\beta=1.0395$) and
$T=T_c/2$ ($\beta=1.386$) both in ferromagnetic regime.

The update scheme involved the Metropolis method applied to over 10\% of the
individual spins, chosen at random, followed by a number (increasing with $L$)
of cluster updates using a Wolff cluster method. See Table~\ref{tab1} for
details.  We refer to each one of these combined updates as a Monte Carlo
sweep.  After every Monte Carlo sweep we measure magnetization and energy,
performing $10^6$ measurements for every pseudosample.

In order to work with thermally equilibrated systems, we performed $10^5$
Monte Carlo sweeps before starting measurements.  We start the simulations
from hot (random) distributions of the spin variables, although we have
checked that the averages do not change if we begin with cold configurations
(i.e., all spins pointing in the same direction).  In Fig.~\ref{thermatest},
we compare the thermalization of the different pseudosamples in the most
challenging case, i.e. our largest system at the lowest temperature.  We
performed a similar check for the log-binning of the specific heat.

\section{Finite-size scaling}
\label{results}	
\setcounter{equation}{0}

\begin{figure}[!t]
\begin{center}
\includegraphics[width=0.34\columnwidth, angle=-90]{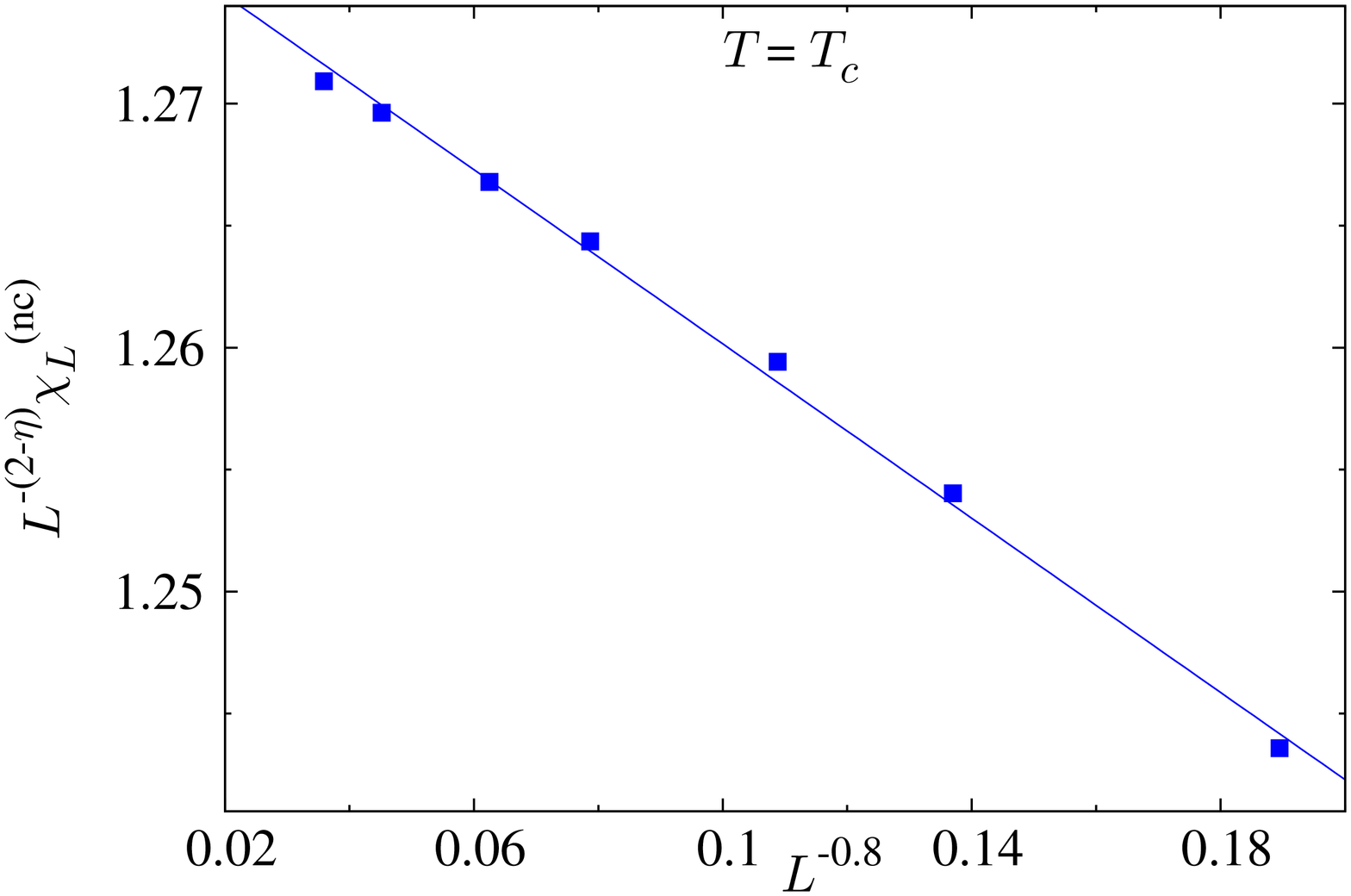}
\includegraphics[width=0.34\columnwidth, angle=-90]{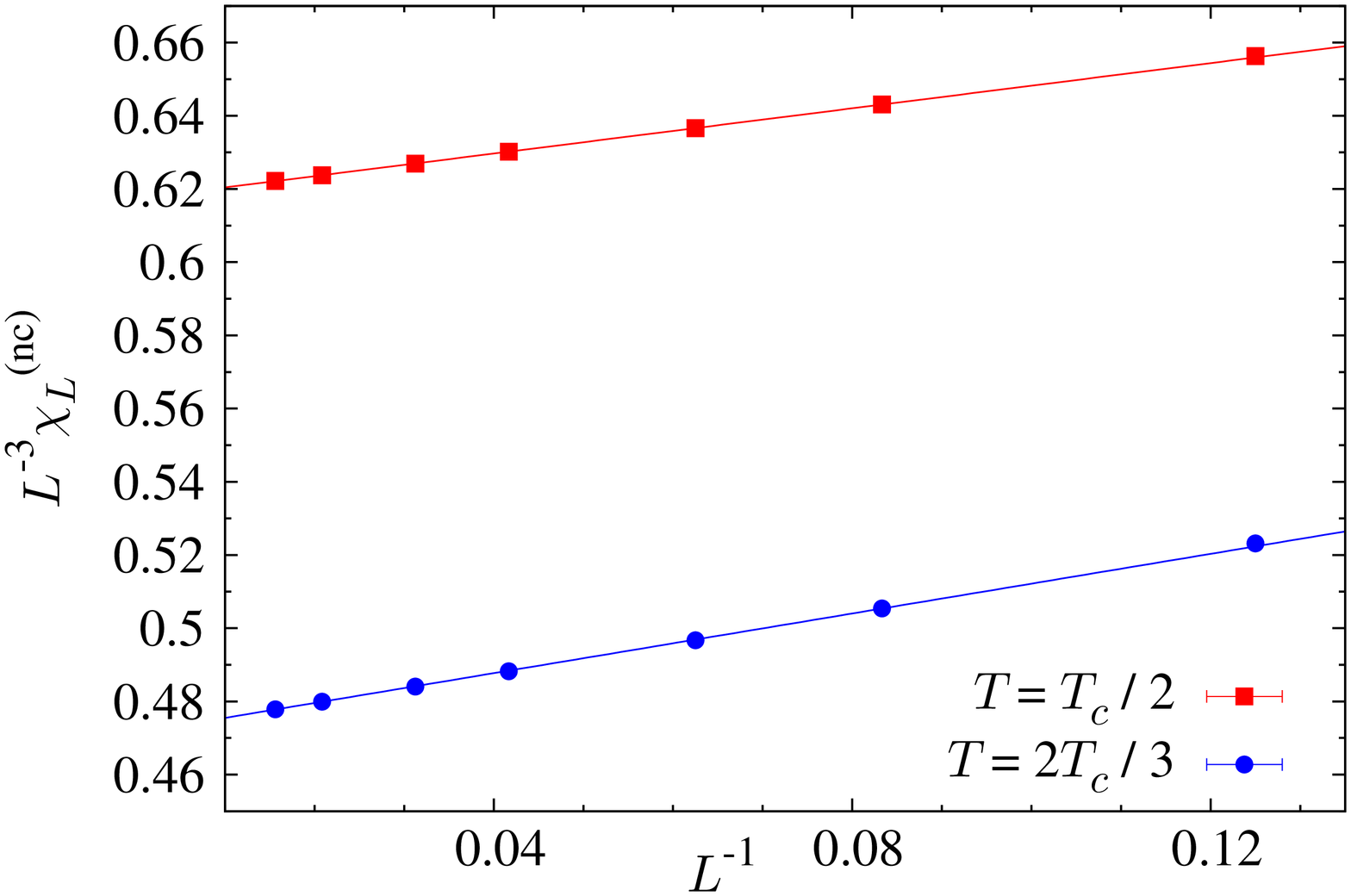}
\caption{Finite-size scaling of the susceptibility at the critical point
  (left) and in the ferromagnetic regime (right), supporting the forms
  (\ref{FSSofchirr}) and (\ref{chibelow}), respectively.  }
\label{fig:susceptibility}
\end{center}
\end{figure}

\begin{table}[!b]
\begin{center}
\begin{tabular}{|r|c|c|c|c|} \hline \hline
$L$  & $r_1(L)$   & $r_2(L)$   & $r_3(L)$   & $r_4(L)$  \\\hline  
8  &   0.00825415(59) &  0.0243727(17)   &  0.0396287(27)   &	 0.0541716(40)  \\
12 &   0.00300815(27) &	 0.00888260(73)  &  0.0144417(12)   &	 0.0197404(21)  \\
16 &   0.00147165(14) &	 0.00434424(34)  &  0.00706250(50)  &	 0.0096533(9)   \\
24 &   0.000537193(39)&  0.00158625(11)  &  0.00257895(18)  &	 0.0035250(3)   \\
32 &   0.000262952(19)&  0.000776455(47) &  0.0012622695(74)& 	 0.00172516(17) \\
48 &   0.0000960985(9)&  0.000283759(24) &  0.0004612664(37)& 	 0.00063038(6)  \\
64 &   0.0000470623(4)&  0.000138962(10) &  0.0002258879(16)& 	 0.00030869(3)   \\
\hline \hline
\end{tabular}
\caption{The first four Lee-Yang zeros for different lattice sizes at $T=T_c$.}
\label{tab2}
\end{center}
\end{table}

We begin our analysis with a brief discussion of FSS of the susceptibility in
the critical and ferromagnetic regimes.  As mentioned in the introduction, our
aim is not to generate new estimates for the critical temperature and critical
exponents.  Rather, we wish to examine previously under-researched aspects of
the Heisenberg model.  Therefore, we first check the consistency of our
results with earlier studies before moving on to the Lee-Yang zeros, which
form the focus for our work.  In Fig.~\ref{fig:susceptibility} (left panel),
the critical susceptibility data are plotted with a best fit to
Eq.(\ref{FSSofchirr}). The estimates $\gamma / \nu = 1.609(9)$ and $\omega =
0.8$ from Ref.~\cite{UexO3_07} are used.  The fit confirms these estimates for
the data.  The susceptibility is also plotted in the ferromagnetic phase in
Fig.~\ref{fig:susceptibility} (right panel).  The scaling form
(\ref{chibelow}) is confirmed, including the corrections coming from the
Goldstone modes.

To achieve relatively small error bars in estimating the zeros, we follow an
iterative approach whereby we first estimate the location of the zeros by
detecting changes in the sign of Eq.(\ref{cos_LY}) using a relatively large
interpolation step and, from this estimation, we restart the search with a
smaller interpolation step.  We terminate this iterative search once the error
bars do not further decrease upon reducing the interpolation step size.  The
estimates for zeros at $T_c$ and below $T_c$ are listed in Tables~\ref{tab2},
\ref{tab4} and~\ref{tab5}.  respectively.

\begin{figure}[!t]
\begin{center}
\vspace{-5cm}
\includegraphics[width=0.7\columnwidth, trim=20 40 0 40, angle=270]{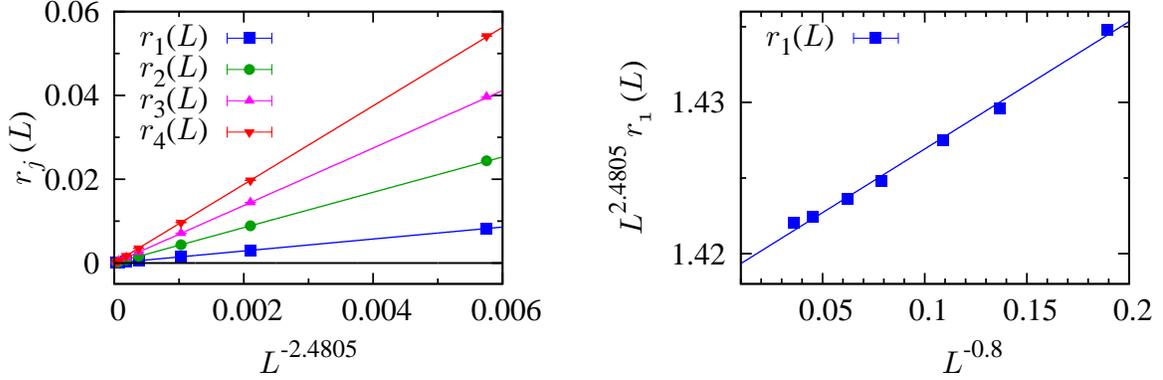}
\caption{Scaling of the first four Lee-Yang zeros at the critical temperature
  (in color online).  The left panel confirms the leading finite-size scaling
  at the critical point as $\chi_L(0) \sim L^{\Delta / \nu} = L^{-2.4805}$
  following Eq.(\ref{FSSofchirr}). The right panel lends support for the
  accepted value of the finite-size correction exponent $\omega = 0.8$.  }
\label{scaling_atTc_zeros} 
\end{center}
\end{figure}

\begin{figure}[!t]
\begin{center}
\includegraphics[width=0.7\columnwidth, angle=270]{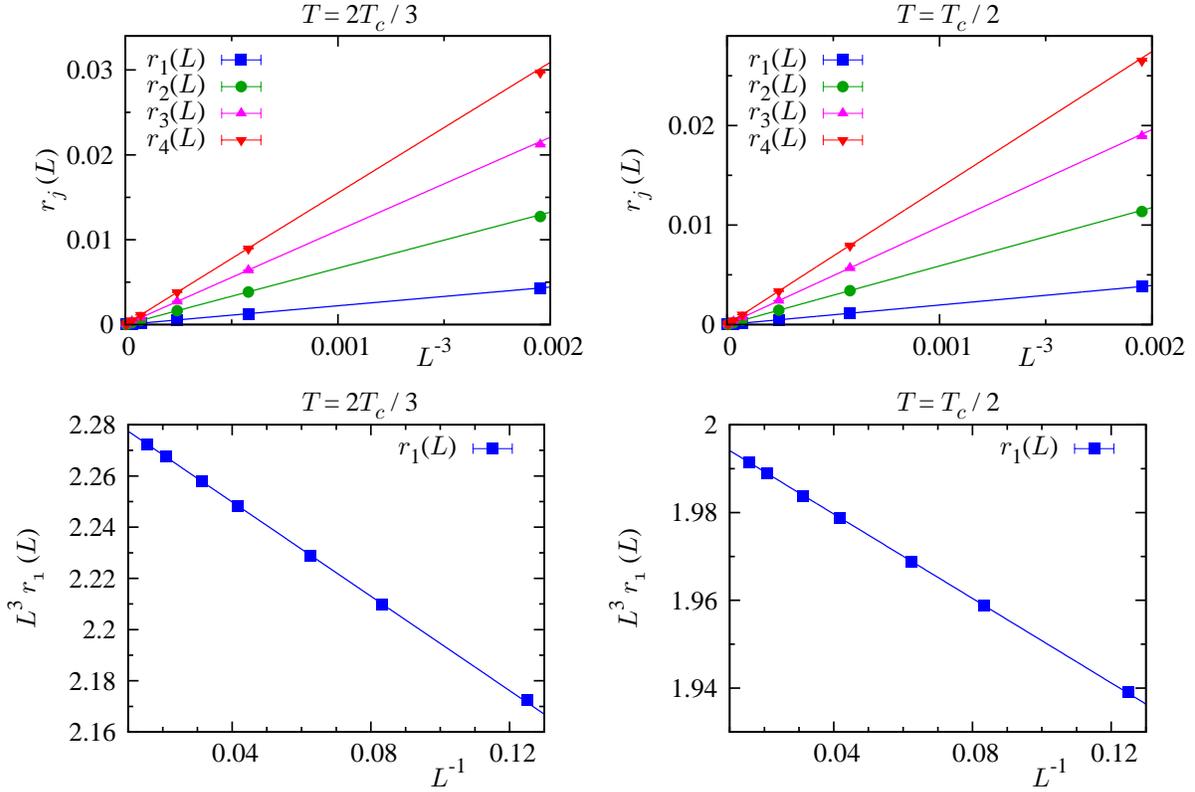}
\caption{Scaling of the first four Lee-Yang zeros below the critical
  temperature (in color online).  The upper panels confirm the leading
  finite-size scaling at the critical point as $\chi_L(0) \sim L^{-d} =
  L^{-3}$. The bottom panels confirm that the associated correction term is
  $L^{-1}$, indicative of the presence of Goldstone bosons.  }
\label{fig_scaling_zeros_belowTc}
\end{center}
\end{figure}

\begin{table}[!b]
\begin{center}
\begin{tabular}{|r|l|l|l|l|} \hline \hline
$L$  & $r_1(L)$   & $r_2(L)$   & $r_3(L)$   & $r_4(L)$  \\\hline  
 8 & 0.00378744(2)     & 0.01136228(5)      & 0.01893702(8)   &  0.0265116(2)    \\
12 & 0.001133597(3)    & 0.003400788(9)     & 0.00566797(2)   &  0.00793514(3)    \\
16 & 0.0004806650(8)   & 0.001441994(2)     & 0.002403321(4)  &  0.003364646(6)     \\
24 & 0.0001431430(2) 	 & 0.0004294297(6)    & 0.000715716(1)  &  0.001002002(2)   \\
32 & 0.00006054220(4)  & 0.0001816267(2)    & 0.0003027111(2) &  0.0004237954(3)  \\
48 & 0.000017984100(9) & 0.00005395236(3)   & 0.00008992060(5)&  0.00012588883(7)   \\
64 & 0.000007596710(4) & 0.00002279013(1)   & 0.00003798356(2)&  0.00005317698(3) \\
\hline \hline
\end{tabular}
\caption{The first four Lee-Yang zeros for different lattice sizes at $T=T_c/2$. }
\label{tab4}
\end{center}
\end{table}

The  scaling dependency of the zeros on the system size is obtained by fitting to
\begin{equation}
r_j(L)=a+bL^{-c}(1+ f L^{-e}).
\label{zeros_scaling_func}
\end{equation}
In the absence of the Yang-Lee edge (i.e., at criticality and in the
ferromagnetic phase), we expect that $a$ should be compatible with zero.  At
the critical point, Eq.(\ref{a77}) predicts that $c = \Delta / \nu$ and $e =
\omega$.  In the ferromagnetic regime, on the other hand, we expect $c = 3$
and $e=1$.  In the paramagnetic region, where the Yang-Lee edge is manifest,
accurate estimates for the zeros should generate a non-vanishing value for
$a$.

\begin{table}[!b]
\begin{center}
\begin{tabular}{|r|l|l|l|l|} \hline \hline
$L$  & $r_1(L)$         & $r_2(L)$          & $r_3(L)$         & $r_4(L)$  \\\hline  
 8 & 0.00424297(3)      & 0.012728664(8)    & 0.0212137(2)     &  0.0296977(2)  \\
12 & 0.001278891(5))    & 0.001632521(8)    & 0.00639437(3)    & 0.00380919(2)  \\
16 & 0.000544175(3)     & 0.0014419935(2)   & 0.00272086(2)    &  0.0033646459(6)     \\
24 & 0.0001626250(4) 	& 0.000487875(1)    & 0.000813124(2)   & 0.001138372(3)   \\
32 & 0.0000689022(1)    & 0.0002067065(3)   & 0.0003445107(5)  &  0.0004823147(7)  \\
48 & 0.00002050330(2)   & 0.00006150994(6)  & 0.0001025166(1)  &  0.0001435232(2)   \\
64 & 0.000008668430(7)  & 0.00002600530(2)) & 0.00004334215(4) & 0.00006067901(5) \\
\hline \hline
\end{tabular}
\caption{The first four Lee-Yang zeros for different lattice sizes at $T=2T_c/3$. }
\label{tab5}
\end{center}
\end{table}

The FSS for the first four zeros at $T=T_c$ using the full magnetization
($\boldsymbol{|M|}$), through solving Eq.(\ref{cos_LY}), is given in
Fig.~\ref{scaling_atTc_zeros}.  A fit to the form~(\ref{zeros_scaling_func})
clearly points to a value $a\approx 0$.  Fixing this value for $a$ and also
fixing $e=\omega=0.8$ leads to the estimates for $\Delta/\nu$ listed in
Table~\ref{tab3}.  All of the estimates are in agreement with the estimate
$\Delta/\nu=2.4805(4)$ coming from Ref.~\cite{UexO3_07}.  Fixing $f=0$, on the
other hand, leads to unacceptable fits.  In this table we also present results
whereby the zeros are obtained using just one of the individual components of
the magnetization vector, in this case $M_x$.  Clearly the scaling results do
not depend on the selection of a specific component.

Next we study the FSS of the zeros below the critical temperature.  The FSS
behavior is plotted in Fig.~\ref{fig_scaling_zeros_belowTc}.  Again we obtain
clear indications that $a=0$, as expected.  Again we do not obtain acceptable
fits for the remaining scaling if we do not include a correction-to-scaling
term.  Fitting for both the leading and sub-leading behavior delivers the
estimates listed Table~\ref{tab6}.  The leading scaling exponent is clearly
equal to 3 in each case, and the correction exponents are very close to 1,
indicating the presence of Goldstone modes, as discussed around
Eq.(\ref{LYGB}).

\begin{table}[!b]
\begin{center}
\begin{tabular}{|r|c|c||c|c|} \hline \hline
 & \multicolumn{4}{|c|}{$T=T_c$}  \\\hline
& \multicolumn{2}{|c||}{$\langle{\cos{(r|\boldsymbol{\mathit{M}}|)}}\rangle = 0$}& \multicolumn{2}{|c|}{$\langle{\cos{(rM_x)}}\rangle = 0$}  \\\hline
 & $c = \Delta/\nu$  & $\chi^2/\mathrm{ndf}$  & $c = \Delta/\nu$   & $\chi^2/\mathrm{ndf}$ \\\hline
$r_1(L)$ &  2.4774(12)   & 0.11 / 2 & 2.4792(7)  & 2.57 / 5  \\\hline
$r_2(L)$ &  2.4789(28)   & 2.52 / 4 & 2.4845(17) & 0.52 / 4   \\\hline
$r_3(L)$ &  2.4791(3)    & 5.36 / 4 & 2.4811(26) & 0.89 / 5  \\\hline
$r_4(L)$ &  2.4793(4)    & 5.48 / 4 & 2.4779(52) & 4.98 / 5   \\\hline
\hline \hline
\end{tabular}
\caption{Scaling exponent of the Lee-Yang zeros measured at the critical
  temperature.  The estimates for the critical exponents are independent of
  the manner in which the zeros were determined, indicated by
  Eq.(\ref{cos_LY}) and Eq.(\ref{cos_LYx}). $\mathrm{ndf}$ is the number of
  degrees of freedom of the fit.}
\label{tab3}
\end{center}
\end{table}

\begin{figure}[!t]
\vspace{-4cm}
\begin{center}
\includegraphics[width=0.7\columnwidth, angle=270]{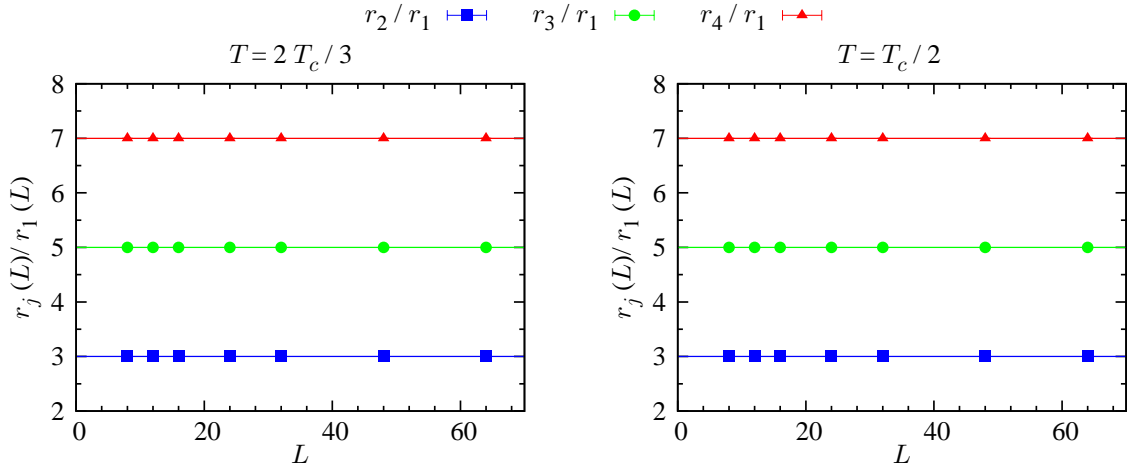}
\caption{The ratio $r_j(L)/r_1(L)$ (in color online) is independent of $L$ in
  the ferromagnetic phase and is, in fact, $2j-1$.}
\label{beta=beta_rj_over_r1_vs_L}
\end{center}
\end{figure}

We also investigate scaling with the index of the zeros, beginning with the
ferromagnetic region.  There, Eq.(\ref{a77}) predicts
\begin{equation}
 \frac{r_j(L)}{r_1(L)}
 =
 \frac{j-\epsilon}{1-\epsilon}.
 \label{4.2}
\end{equation}
This is also investigated in Fig.\ref{beta=beta_rj_over_r1_vs_L} for two values of $T<T_c$.
The two panels clearly indicate that $r_j/r_1$ is independent of $T$ 
and of $L$. Moreover, their numerical values indicate that
\begin{equation}
 \epsilon = \frac{1}{2} \quad {\mbox{for~$T<T_c$.}}
 \label{eps=1o2}
\end{equation}
Therefore the functional form involving the fractional number of zeros,
previously suggested at criticality, extends to the ferromagnetic region too.

\begin{table}[!b]
\begin{center}
\begin{tabular}{|r|c|c|c||c|c|c||} \hline \hline
  & \multicolumn{3}{|c||}{$T=T_c/2$}  & \multicolumn{3}{|c||}{$T=2T_c/3$}  \\\hline 
  & $c$  & $e$  & $\chi^2/\mathrm{ndf}$ & $c$  & $e$  & $\chi^2/\mathrm{ndf}$  \\\hline
$r_1(L)$ &  3.00024(8)  & 0.962(7)  & 2.98 / 2  &  3.0006(2) & 0.949(8)  & 3.88 / 2  \\\hline
$r_2(L)$ &  3.00023(8)  & 0.963(7)  & 2.06 / 2  &  3.0005(2) & 0.952(9)  & 3.68 / 2  \\\hline
$r_3(L)$ &  3.00022(8)  & 0.964(7)  & 1.78 / 2  &  3.0004(2) & 0.956(9)  & 2.92 / 2 \\\hline
$r_4(L)$ &  3.00021(8)  & 0.965(7)  & 1.51 / 2  &  3.0003(2) & 0.962(9)  & 1.84 / 2 \\\hline
\hline
\end{tabular}
\caption{Scaling and correction to scaling exponents, $c$ and $e$ in
  Eq.~(\ref{zeros_scaling_func}), obtained from the Lee-Yang zeros below the
  critical temperature.  The results confirm the prediction $c=d=3$ and $e =
  1$ from Eq.(\ref{a77}). }
\label{tab6}
\end{center}
\end{table}

\begin{figure}[!t]
\begin{center}
\includegraphics[width=0.45\columnwidth, angle=270]{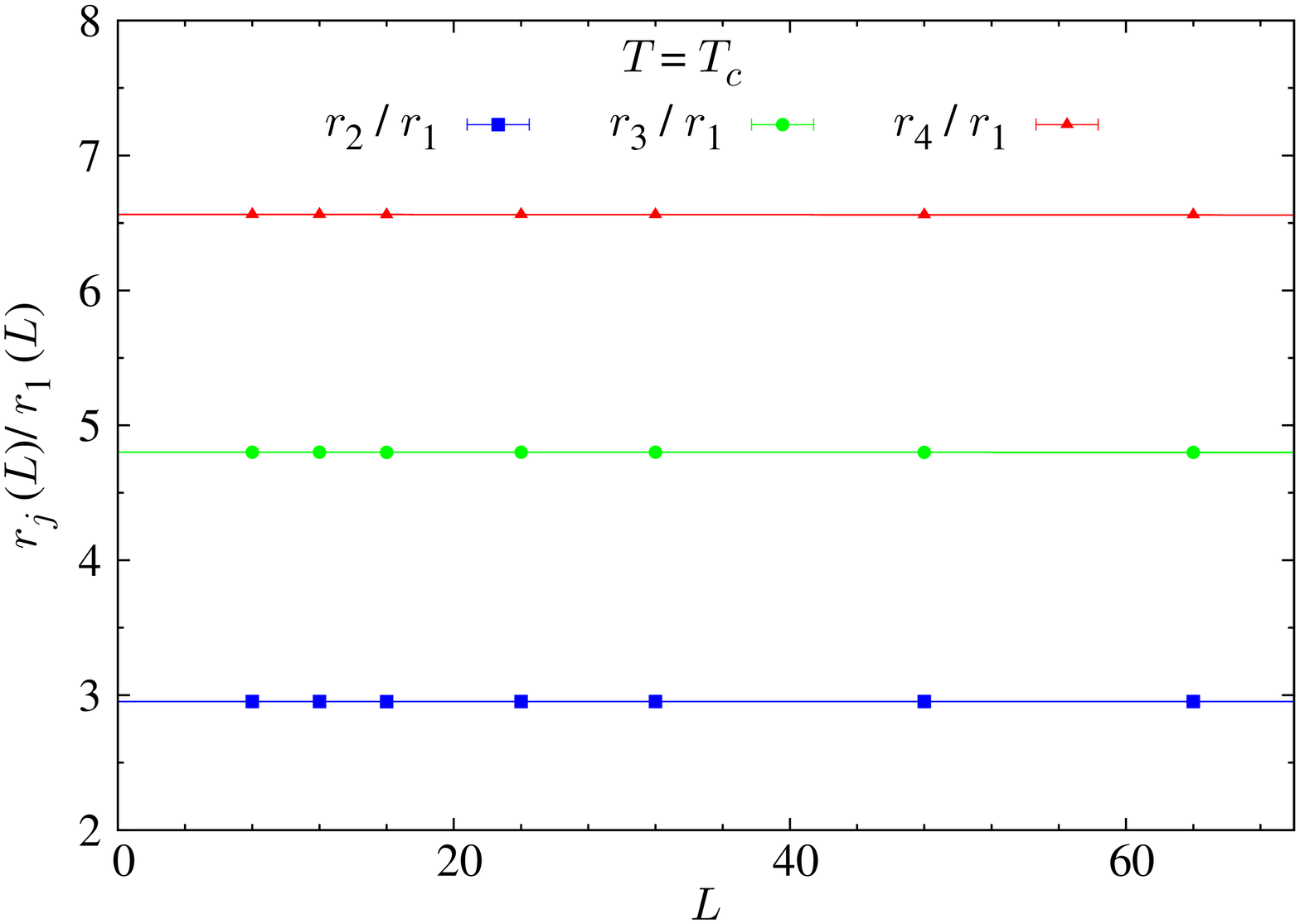}
\caption{The ratio $r_j(L)/r_1(L)$ is independent of $L$ also at the critical point (in color online).   }
\label{epsilon_c}
\end{center}
\end{figure}
The $j$-dependency at the critical point is investigated in
Fig.\ref{epsilon_c}.  One observes that $r_j(L)/r_1(L)$ is also independent of
$L$ at $T_c$.  The values of $r_j(L)/r_1(L)$ are, however, less easy to
interpret than they were in the ferromagnetic case.  The counterpart to
Eq.(\ref{4.2}) is
\begin{equation}
 \frac{r_j(L)}{r_1(L)}
 =
 \left({\frac{j-\epsilon}{1-\epsilon}}\right)^{\frac{\Delta}{\nu d}}
\left\{{1 + {\mathcal{O}}\left({\frac{j-\epsilon}{L^d}}\right)^{\frac{\omega}{d}}}\right\},
 \label{4.22}
\end{equation}
and attempts to extract a precise estimate for $\epsilon$ from this formula
are beset by large errors.  
Indeed, we cannot discount a functional dependency of $\epsilon$ on $j$.
Instead the full dependency may be interpreted in
terms of the density of zeros, and this is analyzed in
Section~\ref{results_density}.

Numerical determination of the locations of the Lee-Yang zeros in the
paramagnetic phase is hampered by considerable limitations in algorithmic
accuracy.  In fact these problems are intrinsically so severe as to yield
spurious zeros and hinder meaningful analysis of the Yang-Lee edge.
For this reason, we relegate the discussion to the Appendix.

\section{Density of zeros}
\setcounter{equation}{0}
\label{results_density}

\begin{table}[b]
\begin{center}
\begin{tabular}{|r|c||c|} \hline 
 $j$ & $a_2$  & $\omega$ \\\hline \hline
1   & 1.2097(4)     & 0.9(3) \\\hline
2   & 1.2094(3)     & 1.1(2) \\\hline
3   & 1.2092(2)     & 1.23(6) \\\hline 
4   & 1.2090(3)     & 1.3(3)  \\\hline 
\hline
\end{tabular}
\caption{$a_2$ and $\omega$ from the density of zeros via Eq.(\ref{eq:exponents_Tc}).}
\label{tab7}
\end{center}
\end{table}

\begin{figure}[!t]
\begin{center}
\includegraphics[width=0.45\columnwidth, trim=0 0 30 0, angle=270]{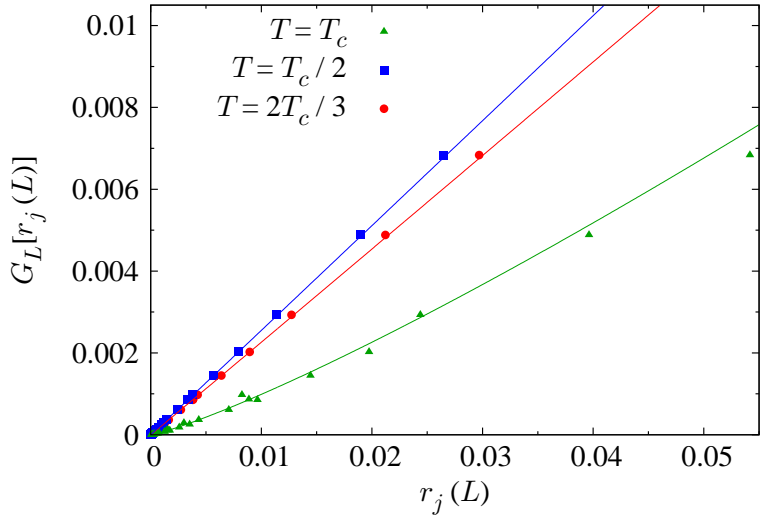}
\caption{The density of zeros for at $T = T_c/2$ (denoted by the symbols $\star$, in blue online), $T=2T_c/3$ ($\times$, green online) and at $T=T_c$ (+, red online). }
\label{density}
\end{center}
\end{figure}

A numerical approach to the determination of the density of partition function
zeros was developed in Refs.\cite{JaKe2001,JaJoKe2004}. The cumulative density
for a finite-size system is defined as
\begin{equation}
 G[r_j(T;L)] = \frac{2j-1}{2L^d}.
 \label{Gdensity1}
\end{equation}
At the infinite-volume critical point $T_c$ this scales in the Lee-Yang case as
\begin{equation}
 G(r) \sim r^{\frac{1}{\delta}+1} = r^{\frac{\nu d}{\Delta}},
 \label{Gdensity2}
 \end{equation}
which is compatible with the compact description of scaling given in
subsection~\ref{compact}.  In the ferromagnetic regime, on the other hand, one
expects the linear behavior~\cite{JaKe2001,JaJoKe2004}
\begin{equation}
 G(r) \sim r.
 \label{Gdensity3}
 \end{equation}
Differentiating Eq.(\ref{Gdensity2}) gives a density of zeros $g(r) \sim
r^{1/\delta}$, commensurate with the magnetic scaling form $ m_\infty(T_c,h)
\sim h^{\frac{1}{\delta}}$.  Differentiating (\ref{Gdensity3}), on the other
hand gives a non-vanishing density of zeros, ensuring a discontinuous
transition across $h=0$.  Here, we wish to test these expectations for the 3D
Heisenberg model.  To do this, we fit our numerical data to the form
\begin{equation}
G_L[r_j(L)]=a_1 [r_j(L)]^{a_2}+a_3 ,
\label{density_scaling_func}
\end{equation}
where the coefficients depend on the temperature.  We employ the fitting
procedure used in Refs.\cite{JaKe2001,JaJoKe2004} whereby, in the absence of
error bars for the density estimates in Eq.(\ref{Gdensity1}), one assumes an
error of $\sigma/L^d$ and then tunes $\sigma$ to deliver a best fit with
chi-squared per degree of freedom of one. This method delivers error estimates
for the fitted parameters but precludes an independent goodness-of-fit test.

\begin{table}[!b]
\begin{center}
\begin{tabular}{|r|l|l||l|l|} \hline \hline
     & \multicolumn{2}{|c||}{$M_{\rm{sp}}$ measured directly}        & \multicolumn{2}{|c|}{$M_{\rm{sp}}$ measured via density}  \\\hline
 $L$ & $T=2T_c/3$      & $T=T_c/2$          & $T=2T_c/3$       & $T=T_c/2$    \\\hline \hline
8    & 0.723070(4)     & 0.810036(3)        & 0.723078(3)      & 0.810038(3)  \\\hline 
12   & 0.710792(3)     & 0.801895(2)        & 0.710794(2)      & 0.801898(2)  \\\hline
16   & 0.704728(3)     & 0.797844(1)        & 0.7047315(20)    & 0.7978451(7) \\\hline 
24   & 0.698713(2)     & 0.793808(1)        & 0.6987122(8)     & 0.7938082(6) \\\hline 
32   & 0.695724(1)     & 0.791793(1)        & 0.6957245(5)     & 0.7917944(3) \\\hline 
64   & 0.691257(1)     & 0.7887771(4)       & 0.6912572(3)     & 0.7887785(2) \\\hline
\hline \hline
\end{tabular}
\caption{Sample averaged spontaneous magnetization below the critical
  temperature measured directly and measured via Eq.(\ref{MG}) below the
  critical temperature using data for each lattice size individually.}
\label{tab8}
\end{center}
\end{table}

The data are plotted Fig.~\ref{density} for $T = T_c/2$, $T=2T_c/3$ and
$T=T_c$.  Fitting to Eq.(\ref{density_scaling_func}) yields $a_3 \approx
10^{-7} \pm 10^{-7}$ in each case.  For the ferromagnetic data, fixing $a_3=0$
and fitting for the remaining parameters delivers $a_2$ compatible with $1$
and supportive of Eq.(\ref{Gdensity3}).  ($a_2= 1.004(1)$ and $a_2=1.007(1)$
for $T=T_c/2$ and $T=2T_c/3$, respectively when all data points are included
in the fits, reducing to $a_2= 1.001(1)$ and $a_2=1.002(1)$ when only the 8
points closest to the origin are used in the fits).  At the critical point
itself, using all data, one estimates $a_2 = 1.203(5)$.  In comparison, the
estimate $\eta = 0.0391(9)$ from Ref.\cite{UexO3_07} delivers $a_2 = \nu
d/\Delta = 2d/(d+2-\eta) = 1.2095(2)$.

While the density plots give reasonable collapse in lattice sizes, we can also
analyze each $L$ independently for greater precision.  In Table~\ref{tab7} we
report the exponents we have obtained assuming a fit, including scaling
corrections, of the form
\begin{equation}
r_j(L)=b_1 G_L^{b_2}\left({ 1+ b_3 G_L^{b_4}  }\right) ,
\label{eq:exponents_Tc}
\end{equation}
using error bars in $r_j(L)$ and not in $G_L$.
With $a_2=1/b_2$ and  
$\omega = d b_4$, we have also obtained reasonable agreement
with  the value $\omega\simeq 0.8$ quoted in the literature \cite{PeVi02}.

Finally, although there is no order parameter for the finite-size system,
according to Lee and Yang's fundamental theory of phase transitions, one can
relate the density of zeros to the value of the spontaneous magnetization and
one expects~\cite{LEEYANG}
\begin{equation}
 M_{\rm{sp}} = \pi a_1 
 .
 \label{MG}
\end{equation} 
We compare measurements of $M_{\rm{sp}}$ via Eq.(\ref{MG}) with direct
estimates of $M_{\rm{sp}} = \langle |\boldsymbol{\mathit{M}}| \rangle$, where
$\boldsymbol{\mathit{M}}$ is defined in Eq.(\ref{vectorialmag}).  We perform
the comparison using the full data sets for each $T<T_c$ as well as for each
lattice size independently.  We have checked that the data for each lattice
follows a straight line according with the theoretical expectation and present
our results in Table~\ref{tab8}.  We list in Table~\ref{tab8} the different
estimates of the spontaneous magnetization for $T<T_c$ for the different
lattice sizes. The agreement between them is excellent.

\section{Conclusions}
\label{Con}
\setcounter{equation}{0}

We have performed a numerical analysis of the Heisenberg model in three
dimensions, paying special attention to the Lee-Yang zeros, the scaling
properties of which contain information on Goldstone modes.  Besides FSS for
individual zeros in the critical and paramagnetic regimes, we have looked at
the index of zeros and shown that a comprehensive description extends to both
regions.  These allow us to confirm very precise estimates for the critical
exponents and correction terms.  A first attempt to numerically examine
scaling associated with the Yang-Lee edge in the paramagnetic region
encounters obstacles which we elucidate directly and through analogy with the
1D Ising model.  Finally, we confirm that study of the density of zeros for
finite size, offers a compact way to investigate the onset of spontaneous
magnetization, although the latter is only manifest in infinite volume.

\vspace{1cm}
\noindent
{\bf{Acknowledgments:}} This research was supported by Marie Curie
International Incoming Fellowship and International Research Staff Exchange
Scheme grants within the 7th European Community Framework Program.  AGG and
JJR acknowledge support from Research Contracts No. FIS2007-60977 (MICINN),
GR10158 (Junta de Extremadura) and PIRSES-GA-2011-295302 (European Union).  RK
thanks Nickolay Izmailian for discussions.

%

\appendix

\section{spurious zeros in the paramagnetic phase}
\label{App}
\setcounter{equation}{0}

\begin{figure}[!t]
\begin{center}
\includegraphics[width=0.5\columnwidth, angle=270]{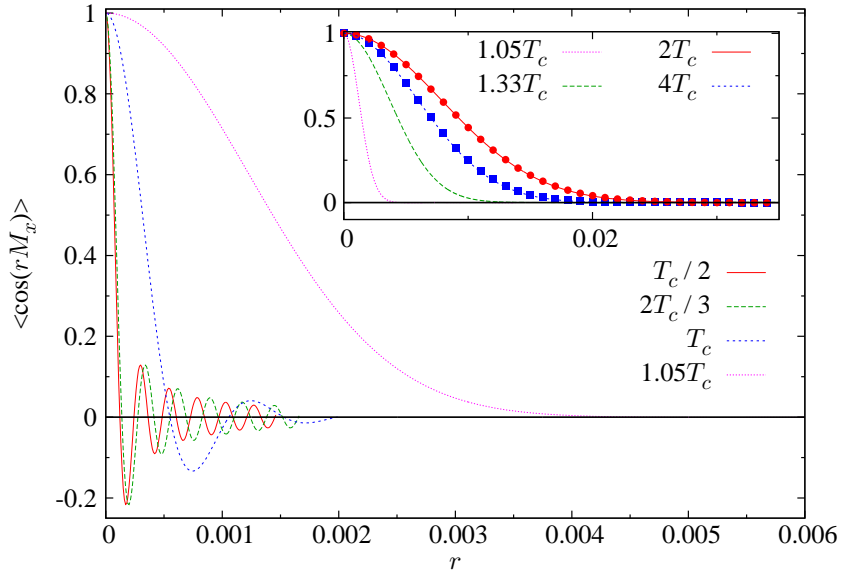}
\caption{Behavior of $\langle \cos{(rM_x)} \rangle$ for the 3D Heisenberg
  model with $L=32$ below, at, and above criticality (in color online). We
  have also plotted in the inset the behavior in the paramagnetic phase in
  order to show up the very different behavior there. In addition, we have
  plotted for the two highest temperatures the prediction from the Gaussian
  approximation (continuous lines) , see Eq. (\ref{false3}), and the points
  (red and blue)
  from the numerical simulations: there are no free parameters, since
  we have used the susceptibility which was computed numerically. Notice the
  good agreement. For $T=1.05 T_c$ and $T=1.33 T_c$ we only show the data from
  numerical simulations (dashed lines).}
\label{cosine_evolution}
\end{center}
\end{figure}

Fig.~\ref{cosine_evolution} shows the evolution of the expectation of the
cosine in Eq.(\ref{cos_LYx}), through which the zeros are detected.  One
notices a remarkable difference between the amplitudes of the function below
and above criticality; in the paramagnetic phase the amplitude of $\langle
\cos(r M_x) \rangle$ is dampened as $r$ increases, an effect not present at or
below criticality.  This leads to algorithmic detection of spurious Lee-Yang
zeros in the symmetric phase.

That the detected zeros are indeed spurious is indicated firstly by a
straightforward fit to Eq.(\ref{zeros_scaling_func}), which delivers $a
\approx 0$.  The fact that the estimated zeros do not settle onto a Yang-Lee
edge already hints that they are spurious.  A second feature is that the
scaling appears to indicate a leading exponent $c \approx 1.5 = d/2$.  That
this is also spurious is indicated as follows.

It is well known that the probability distribution of the magnetization in the
paramagnetic phase follows an approximate Gaussian probability distribution.
We write this distribution as (considering a single dimension here for simplicity)
\begin{equation}
P(M)=\frac{1}{ \sqrt{2 \pi V \chi_L^{\rm{(nc)}}}} \exp\left[ -\frac{M^2}{2 \chi_L^{\rm{(nc)}} V} \right],
\end{equation}
where $M$ is the total magnetization, $V$ is the volume and
$\chi_L^{\rm{(nc)}}=\langle M^2 \rangle /V$ is the susceptibility, which is
finite in the paramagnetic phase.  The algorithm detects zeros through
Eq.(\ref{cos_LYx}), and with the Gaussian distribution governing the
high-temperature phase,
\begin{equation}
\langle \cos(r M) \rangle = \exp\left[ -\frac{1}{2} \chi_L^{\rm{(nc)}} V r^2 \right]
\label{false}
\end{equation}
there. Here we have assumed  $V \gg 1$ (otherwise this result would be modulated
by an error-function factor). 
Therefore $\langle\cos(r M) \rangle$ decays exponentially quickly in the
paramagnetic phase.\footnote{In $O(N)$ models one obtains:
\begin{equation}
\langle \cos(r M_x) \rangle = \exp\left[ -\frac{1}{2 N} \chi_L^{\rm{(nc)}} V r^2
  \right] \,.
\label{false3}
\end{equation}
} The reader can see the suitability of the Gaussian approximation in the inset
of Fig.~\ref{cosine_evolution}.

Numerically we compute $c(r)=\langle \cos(r M) \rangle$ with a given
statistical error (which is also $r$-dependent) that we will denote
$\delta(r)$. When $c(r_*)\sim \delta(r_*)$, a statistical fluctuation can
induce a spurious zero at $r_*$. Hence, if we have similar error bars for all
the lattice sizes, this implies, see Eq.(\ref{false}), that the spurious zero
scales as $1/\sqrt{V}$.  This explains the origin and scaling of the spurious
paramagnetic zeros -- The behavior is simply due to finite statistics
associated with the numerical approach.  Instead, if we improve the
statistics, reducing the value of $\delta(r_*)$, the spurious zero should be
disappear.

We can gain further insight by examining slope of $\langle \cos(r M) \rangle$,
which is
\begin{equation}
\frac{d}{d r}\langle \cos(r M) \rangle = -\langle M \sin(r M) \rangle .
\end{equation}
We can examine this slope in the three different regimes.  At and below the
critical point, we use the fact that $M \simeq \sqrt{\langle M^2\rangle}$, to
see that in both cases $r_1 M$ is $O(1)$, where $r_1$ is a genuine zero (and
having used the scaling of the zeros in each of these two regions). Therefore,
close to $r_1$,
\begin{equation}
\left.\frac{d}{d r}\langle \cos(r M) \rangle\right|_{r_1} \sim  |M|   .
\end{equation}
Since $|M| \sim V$ in the ferromagnetic phase, and its typical value at
criticality is $|M| \simeq \sqrt{\langle M^2\rangle}=\sqrt{V
  \chi_L^{\rm{(nc)}}}$, it is clear that at and below the critical temperature
the slope is large. Since the algorithm detects zeros through changes in the
sign of $\langle \cos(r M) \rangle$, it is robust in the critical and
ferromagnetic regions.  In the paramagnetic region, however, the Gaussian
approximation gives
\begin{equation}
\left.\frac{d}{d r}\langle \cos(r M) \rangle\right|_{r_1}= -r_1 V \chi 
\exp\left[ -\frac{1}{2} \chi V r_1^2
  \right].
\end{equation}
This gives an exponentially depressed slope in the paramagnetic phase,
rendering detection of genuine zeros difficult and spurious zeros (as noise)
feasible.

\begin{figure}[!t]
\begin{center}
\includegraphics[width=0.5\columnwidth, angle=270]{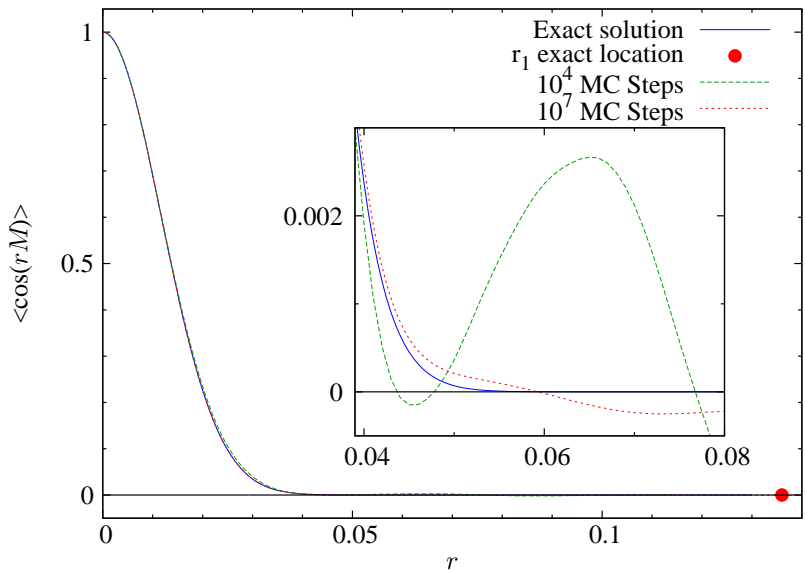}
\caption{Behavior of $\langle \cos{(rM)} \rangle$ for the 1D Ising model
  with $L= 1000$ above criticality, i.e., above $T=0$ (in color online). The
  exact zero is indicated by the red disc and the exact soluction by a blue
  line. The numerical algorithm, however, detects spurious zeros as can be
  seen in the inset.  }
\label{cosineevolution}
\end{center}
\end{figure}

To check the above interpretation, we refer to the Ising model in one
dimension, where the partition function in a magnetic field can be
analytically determined using periodic boundary conditions and where the
entire $T>0$ region is paramagnetic \cite{LEEYANG}.  The two eigenvalues of
the transfer matrix are
\begin{equation}
\lambda_\pm(\beta,H)=e^\beta \left[\cosh(H) \pm \sqrt{e^{-4 \beta} + \sinh^2(H)}\right],
\end{equation}
and the partition function of a chain of $L$ spins is
\begin{equation}
Z(\beta,H)=\lambda_{+}(\beta,H)^L + \lambda_{-}(\beta,H)^L.
\end{equation}
Introducing a pure imaginary magnetic field by defining
$H=ir$, the eigenvalues can be written
\begin{equation}
\lambda_\pm(\beta,ir)=e^\beta \left[\cos(r) \pm \sqrt{ e^{-4 \beta} -\sin^2(r)}\right].
\end{equation}
Notice that for $e^{-2 \beta} < \sin^2(r)$, the eigenvalues, $\lambda_\pm$,
are complex numbers but satisfying $\lambda_{+}^*=\lambda_{-}$.  This confirms
our earlier statement that the partition function in a pure imaginary magnetic
field is real.  One finds \cite{KuFi79}
\begin{equation}
\langle\cos(r M) \rangle=\frac{Z(\beta,ir)}{Z(\beta,0)}=\frac{\lambda_{+}(\beta,i r)^L
  +\lambda_{-}(\beta,i r)^L}{\lambda_{+}(\beta,0)^L + \lambda_{-}(\beta,0)^L}.
\end{equation}
Therefore the zeros in the paramagnetic phase of the one-dimensional Ising
model can be exactly determined.  In Fig.~\ref{cosineevolution}, the first
zero for such a system is depicted as a disc (red online).  This figure also
depicts the results for $\langle{\cos{(rM)}}\rangle$ from two Monte Carlo
simulations and for the exact solution.  As expected, the numerically computed
$\langle{\cos{(rM)}}\rangle$ decays rapidly with increasing $r$, it then
remains very close to zero and traverses the axis well before the true zero is
reached (green line).  Although the situation improves with increased
numerical accuracy (see red line), the figure clearly demonstrates that the
Lee-Yang edge is not reliably accessible using this numerical technique.


\begin{thebibliography}{99}



\bibitem{O03}
M.N. Barber, in Phase Transitions and Critical Phenomena, 
edited by C. Domb and J.L. Lebowitz (Academic, New York, 1983), Vol. 8.

\bibitem{PhFr10}
M.H.~Phan, V.~Francoa, N.S.~Bingham, H.~Srikanth, N.H.~Hur, S.C.~Yu,
Journal of Alloys and Compounds {\bf{508}} (2010) 238–244.


\bibitem{DhDh12}
N.~Dhahri, J.~Dhahri, E.K.~Hlil and  E.~Dhahri
Journal of Magnetism and Magnetic Materials {\bf{324}} (2012) 806.



\bibitem{CaHa02}
M.~Campostrini, M.~Hasenbusch, A.~Pelissetto, P.~Rossi and E.~Vicari,
Phys. Rev. B {\bf{65}}(2002) 144520.


\bibitem{UexO3_07}
A.~Gordillo-Guerrero and J.J.~Ruiz-Lorenzo, J. Stat. Mech. {\bf P06014} (2007).




 \bibitem{PeVi02}
A. Pelissetto and E. Vicari, Phys. Rept. {\bf 368} (2002) 549 .


\bibitem{LEEYANG} T.D. Lee and C.N. Yang, Phys. Rev. Lett. {\bf 87} (1952) 404; 
ibid {\bf 87} (1952) 410 


\bibitem{Wu}
F.Y.~Wu, Talk given at Symposium in honor of Professor C. N. Yang's 85th birthday, Nanyang Technological University, Singapore, November 2007 (arXiv:1010.1838).




\bibitem{GaMSRo67}
G. Gallavotti, S. Miracle--Sole and D.W. Robinson,
Phys. Lett.A {\bf{25}} (1967) 493;
Commun. Math. Phys.
{\bf{10}} (1968) 311.




\bibitem{Fi78}
M.E. Fisher,
Phys. Rev. Lett. {\bf{40}} (1978) 1610.


\bibitem{Asano}
T.~Asano, Phys. Rev. Lett. {\bf{24}} (1970) 1409.



\bibitem{JJRu97}	
J.~J.~Ruiz-Lorenzo,  J. Phys. A {\bf 30}, 485 (1997).


\bibitem{IPZ83} C. Itzykson, R.B. Pearson and J.B. Zuber, Nucl. Phys. B {\bf{220}} (1983) 415.

\bibitem{JaKe2001} W.~Janke and R.~Kenna, J. Stat. Phys. {\bf 102}, 1211 (2001).

\bibitem{Janus} 
R.A.~Ba{\~{n}}os, J.M.~Gil-Narvion, J.~Monforte-Garcia, 
J.J.~Ruiz-Lorenzo and D.~Yllanes, 
 J. Stat. Mech. (2013) P02031.

\bibitem{UCMOND3}
H.~G.~Ballesteros, L.~A.~Fern\'andez, V.~Mart\'{\i}n-Mayor, and A.~Mu\~noz-Sudupe, Phys.~Lett.~B {\bf 387} (1996) 125. 


\bibitem{KuFi79}
D.A. Kurtze and M.E. Fisher,
J. Stat. Phys.
{\bf{19}} (1978) 205;
Phys. Rev. B
{\bf{20}} (1979) 2785.



\bibitem{JaJoKe2004} 
W.~Janke, D.A.~Johnston and R.~Kenna, 
Nucl. Phys. B 682 (2004) 618.


 

\end{thebibliography}
\end{document}